\documentclass [12pt]{article}
\usepackage{graphicx}

\begin{document}

\noindent
{\bf Open chemical reaction networks, steady-state loads and 
Braess-like paradox}

\vskip .1in

Kinshuk Banerjee and Kamal Bhattacharyya\footnote{Corresponding 
author; e-mail: pchemkb@gmail.com}

\vskip .05in
{\it Department of Chemistry, University of Calcutta, 
92 A.P.C. Road,

Kolkata 700 009, India.}

\begin{abstract}
Open chemical reaction systems involve matter-exchange 
with the surroundings. As a result, species can accumulate 
inside a system during the course of the reaction. 
We study the role of network topology in governing the 
concentration build-up inside 
a fixed reaction volume at steady state, particularly 
focusing on the effect of additional paths. 
The problem is akin to that in traffic networks where 
an extra route, surprisingly, can {\it increase} the overall 
travel time. This is known as the Braess' paradox. 
Here, we report chemical analogues of such a paradox 
in suitably chosen reaction networks, 
where extra reaction step(s) can {\it inflate} the 
total concentration, denoted as `load', at steady state. 
It is shown that, such counter-intuitive 
behavior emerges in a qualitatively similar pattern 
in networks of varying complexities. 
We then explore how such extra routes affect the load in 
a biochemical scheme of 
uric acid degradation. From a thorough analysis of this 
network, we propose a functional role 
of some decomposition steps that can trim the load, 
indicating the importance of the latter in 
the evolution of reaction mechanisms in living matter. 
\end{abstract}



\section{Introduction}

Open chemical systems remain an active field of 
research, particularly due to their link with 
a broad class of areas starting from 
living organisms to industrial processes
\cite{Bert,Horn,Fein,Clr,Schus,Bhal,Wagn}. 
Emergence of features like sustained oscillation \cite{Gold} and 
multistability \cite{Edel} in open chemical reaction networks (CRNs) 
\cite{Fein1}
have intimate relations with physiological functions 
like circadian rhythms \cite{Mer} and cellular differentiation 
\cite{Kelr}. 
Complex CRNs exchanging matter (and energy) with 
the environment are of prime importance in biology, {\it e.g.,} 
metabolic \cite{Jeo}, neural \cite{Brw}, gene \cite{Fran} and 
protein networks \cite{Slu}. 
Comparatively small-scale CRNs are ubiquitous in 
the form of tautomeric equilibria, 
important in many biological processes, {\it e.g.}, 
in the chemical versatility of thiamin \cite{Meyr} and 
in spontaneous mutation \cite{Jac}.

Open systems involve flow of material, even in the 
steady state (SS), and 
necessarily describe out-of-equilibrium situations 
\cite{Katch,Prig3,Mou}. 
So, they continue 
to play a fundamental role in the development of non-equilibrium 
thermodynamics \cite{Ross,gasprd1,gasprd,seif1} and, in turn, to understand the 
effects of such far-from-equilibrium scenarios on system 
functionality \cite{gasprd2,Qan1,Qan2}. 
The major points of departure in the behavior of an open 
system compared to a closed system are the following: 
(i) Open systems can support a (non-equilibrium) SS whereas closed systems attain thermodynamic equilibrium \cite{Qan1}; 
(ii) The SS composition depends on the dynamics, {\it i.e.}, on the 
values of the dynamical parameters (rate constants and matter flows); 
but, in an equilibrium state, the concentrations are constrained by  thermodynamics, {\it i.e.}, by equilibrium constants (ratios of rate constants). 
The wealth of behaviors emerging in such (non-equilibrium) SS have generated, over the years, a lot of attention regarding its role in biochemical environments residing far-from-equilibrium. 
Some notable applications are kinetic proofreading \cite{Hop}, 
energy transduction \cite{TH}, optimization of reaction yield 
\cite{Jul}, information acquisition in DNA replication \cite{gaspnas} 
etc.

The theoretical modeling of open CRNs is often based on the 
chemiostatic condition \cite{Qan3,Min,Banj} that still holds a 
key place 
in the irreversible thermodynamic description 
of such networks \cite{Pol}. 
As the name suggests, here concentrations 
of some of the reacting species are taken time-independent 
throughout the course of the reaction. 
In many instances, this leads to 
a simplification of the problem by converting certain 
non-linear kinetic equations to linear ones \cite{Qan1,Qan3,Banj}. 
In this type of situation, however, the matter-exchange with 
the surroundings 
is {\it not} taken into account directly in the formulation of 
rate equations. 
Here, though, we treat the kinetics by explicitly including 
the matter-flow.

In an open CRN, the build-up of reacting species depends on the system  parameters. The latter include the rates of inflow, outflow and 
the reaction constants. 
Here, we denote the total accumulated material inside the 
(fixed) reaction volume as the `load'. 
It is the sum of concentrations of all the species in the reaction 
medium. 
In a closed system, with mass-conservation being valid, 
the load remains the same throughout the 
course of the reaction. So, the concept of load is 
relevant only for open chemical systems where mass-conservation is 
broken. 
However, it seems that such a quantity did not attract much attention. 
The magnitude of the load is important 
because a high value of a component may lead to undesirable 
side-reactions 
such as aggregation; the effects of activity coefficients 
get pronounced as well. 
Also, for a given SS flow, there 
must be an upper limit of the load which the reaction volume can 
sustain. Hence, it is important to understand 
the roles of the network structure and system parameters 
in governing the SS load. This knowledge can then be 
utilized to formulate strategies to reduce the load by 
properly modifying the network for an optimized flow \cite{Tim}. 
This is also crucial in traffic networks where one can take the total 
number of vehicles in various terminuses or traffic signals 
as the load. However, the plan of adding new paths (roads) 
to trim the load and/or improve the flow does {\it not} 
succeed always. In the context of traffic flow, addition 
of a road of much better quality may lead to an {\it increase} in 
the overall travel time from a given source to destination 
for each traveller. 
This is known as the Braess' paradox (BP) \cite{Bra,Rou}, 
after Dietrich Braess, 
who introduced this concept regarding traffic assignment 
problems. The BP also occurs in electrical, mechanical and thermal  networks \cite{Pench,Tim1}. 
For example, BP is manifested in the conductance drop 
of a branched semiconductor 
mesoscopic network when an extra branch is added \cite{Pala}. 
A recent study also reports BP in a chemical system where 
an additional reaction step results in the lowering of 
product-formation rate \cite{Lep}.

\begin{figure}[tbh]
\centering
\rotatebox{270}{
\includegraphics[width=7cm,keepaspectratio]{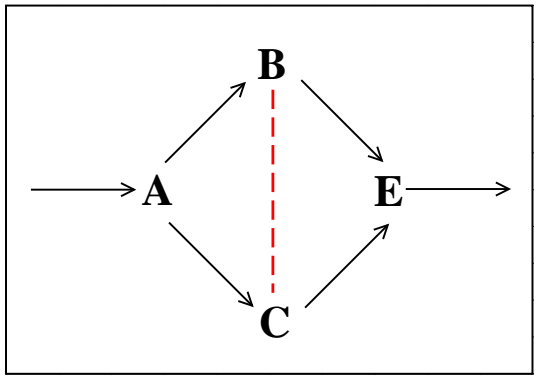}}
\caption{Schematic diagram of an open network. The 
presence of the extra path (dashed, red line), 
which can be uni- or bi-directional, is expected to 
reduce the load at steady state. However, this may not {\it 
always} be the case, leading to a Braess-like paradox.}
\label{fig0}
\end{figure}
In this work, we consider an open CRN, starting 
from a basic four-node structure as shown in Fig.\ref{fig0}. 
The dashed (red) line BC in Fig.\ref{fig0} represents the 
extra path which can be uni- or bi-directional. 
Specific chemical examples of such networks are 
tautomeric (and conformational) equilibria of hydrazones derived from hydroxyl naphthaldehydes where the extra path is absent \cite{Mart} and 
tautomeric equilibria of 
6-(2-pyrrolyl)pyridazines where the extra path is present \cite{Jon}. 
Keto-enol and imine-enamine tautomerism of 
2-, 3- and 4-Phenacylpyridines \cite{Car} and 
imine-enamine tautomerism of the enzyme glyoxylate carboligase 
\cite{Nem} also 
involve four-node networks or subnetworks with or without the extra path. 
Our goal here is to study the effect of 
the extra path on the load, {\it i.e.,} the total concentration of 
the reacting species, at SS. The presence of the extra path 
is expected to decrease the load as it opens up a new avenue 
of transport. 
Thus, a rise in the load when the extra path 
is operative can be designated as a BP-like behavior. 
In Sec.2, we first study a CRN similar to that shown 
in Fig.\ref{fig0} where, except the extra path, all the 
other steps are irreversible. In Sec.3, some of the 
steps are made reversible. The network is extended 
to a six-node one in Sec.4. In all the cases, we find that 
BP-like features appear with a qualitatively similar pattern. 
In Sec.5, the methodology is applied to a CRN of biological 
relevance. 
The overall outcomes are finally summarized in Sec.6.

\section{A network with irreversible edges}

\begin{figure}[tbh]
\centering
\rotatebox{270}{
\includegraphics[width=7cm,keepaspectratio]{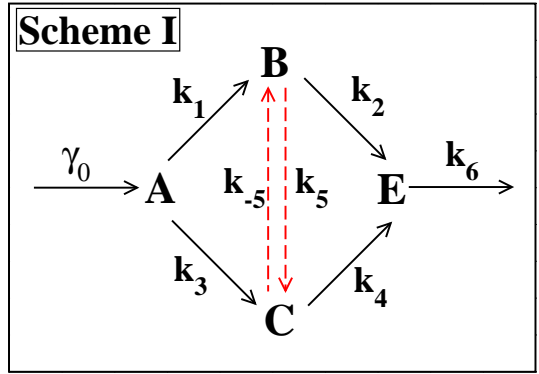}}
\caption{Schematic diagram of a CRN with irreversible edges. 
The extra path is shown by the dashed, red line.}
\label{sch1}
\end{figure}

A simple CRN is shown in Fig.\ref{sch1}. 
We call this network Scheme I. 
Here, $\gamma_0$ is the {\it constant} rate of injection of 
species A into the reaction system of fixed volume and $k_i$ are 
the (first-order) rate constants. Species E goes out of 
the system via a process having rate constant $k_6$. 
The time-dependent concentrations of species A, B, C, E are 
denoted by $a(t),b(t),c(t),e(t)$, respectively. 
Thus, the load $Z(t)$ of this system is
\begin{equation}
Z(t)=a(t)+b(t)+c(t)+e(t). 
\label{zt}
\end{equation}
The thermodynamic fluxes\cite{Prig3} $J_i\,(i=0,1,\cdots,6)$ 
are given by
$$J_0=\gamma_0,\,J_1(t)=k_1a(t),\,J_2(t)=k_2b(t),\,J_3(t)=k_3a(t),$$
\begin{equation}
J_4(t)=k_4b(t),\,J_5(t)=k_5b(t)-k_{-5}c(t),\,J_6(t)=k_6e(t).
\label{J1}
\end{equation}
The reason for introducing these fluxes will become clear later. 
Now, the kinetic equations can be expressed in terms of $J_i$ 
as follows
\begin{equation}
\dot{a}(t)=J_0-J_1(t)-J_3(t)
\label{a}
\end{equation}
\begin{equation}
\dot{b}(t)=J_1(t)-J_2(t)-J_5(t)
\label{b}
\end{equation}
\begin{equation}
\dot{c}(t)=J_3(t)+J_5(t)-J_4(t)
\label{c}
\end{equation}
\begin{equation}
\dot{e}(t)=J_2(t)+J_4(t)-J_6(t).
\label{e}
\end{equation}
At steady state (SS), we have $\dot{a}=\dot{b}=\dot{c}=\dot{e}=0$ and 
hence, $J_0=J_6$ (the SS values are represented by 
omitting the time-argument). From Eqs.(\ref{J1})-(\ref{e}), 
one gets the SS concentrations as
\begin{equation}
a=\gamma_0/(k_1+k_3)
\label{ass}
\end{equation}
\begin{equation}
b=\frac{\gamma_0(k_4+k_{-5})-ak_3k_4}{k_2(k_4+k_{-5})+k_4k_5}
\label{bss}
\end{equation}
\begin{equation}
c=\frac{\gamma_0k_{5}+ak_2k_3}{k_2(k_4+k_{-5})+k_4k_5}
\label{css}
\end{equation}
\begin{equation}
e=\gamma_0/k_6.
\label{ess}
\end{equation}
It is important to note that, the SS concentrations do {\it not} depend 
on the initial condition, {\it i.e.}, the set of concentrations at $t=0$. 
In what follows, we mainly concentrate on the load Z defined in 
Eq.(\ref{zt}) at the SS.

\subsection{Effect of extra path on the load}

Our focus now is on the reaction path BC with 
rate constants $k_{\pm 5}$ that acts as the extra path. 
It is expected that, presence of this extra path should 
enhance the transport efficiency and hence decrease the 
load $Z$ of the system at SS. 
To measure the effect of extra path on $Z$, we introduce the 
load-difference $\Delta$ defined as 
\begin{equation}
\Delta=Z^0-Z
\label{del}
\end{equation}
where $Z^0$ is the value of $Z$ when $k_{\pm5}=0$. 
For finite non-zero values of $k_5$ and/or $k_{-5}$, 
$\Delta$ is expected to be positive. 
We will investigate whether 
this is {\it always} the case. 
If, for a set of parameter values, $\Delta$ comes out to be 
negative (or zero), then 
it is a counter-intuitive behavior analogous to the BP. 
We call any such region of parameter space with 
$\Delta\le0$ as the BP zone. 
The SS concentrations at 
$k_{\pm5}=0$ are similarly denoted by a zero superscript with 
$a^0=a,\,e^0=e$.

Using Eqs.(\ref{bss})-(\ref{css}), $J_5$, the flux of the extra path, 
at the SS reads as
\begin{equation}
J_5=\frac{\gamma_0(k_1k_4k_5-k_2k_3k_{-5})}{(k_1+k_3)
(k_2(k_4+k_{-5})+k_4k_5)}.
\label{J5ss}
\end{equation}
$J_5>0$ means the (net) flow in the extra path is in the direction 
${\rm B\to C}$ and $J_5<0$ indicates that the flow is reversed. 
Now, using Eqs.(\ref{ass})-(\ref{J5ss}), we can 
relate $\Delta$ with $J_5$ by
\begin{equation}
\Delta=\left(b^0+c^0\right)-(b+c)
=\left(\frac{k_4-k_2}{k_2k_4}\right)J_5.
\label{del1}
\end{equation}

\subsection{Emergence of BP zone: finite and infinite}

From Eq.(\ref{J5ss}) and Eq.(\ref{del1}), 
one can see that $\Delta$, 
viewed as a function of $k_4$, becomes zero at 
\begin{equation}
k_4=\left(\frac{k_2k_3k_{-5}}{k_1k_5}\right)\,\,\,{\rm and}\,\,k_4=k_2.
\label{del0}
\end{equation}
$J_5$ is zero at the first solution in Eq.(\ref{del0}). 
So $\Delta$ {\it changes sign twice} as a function of $k_4$. 
When $k_4$ is much smaller than the other rate constants, 
it follows from Eqs.(\ref{J5ss})-(\ref{del1}) that $\Delta$ 
is large and positive. As $k_4$ increases, $\Delta$ 
decreases and becomes zero at the first or the second 
value of $k_4$ given in Eq.(\ref{del0}), 
depending on whether $k_3k_{-5}<k_1k_5$ or not. 
Between these two values, $\Delta$ remains {\it negative}. 
Thus, the extra path leads to an {\it increase} in load 
over {\it a finite region} of parameter space. We call 
this region the {\it finite} BP zone. 
It is easy to check that $\Delta$ shows a similar behavior as 
a function of $k_2$. However, as a function of any other rate 
constant, $\Delta$ changes sign {\it only} once and, 
depending on the parameter values, it can remain negative 
for a large range, approaching zero asymptotically. 
This results in an {\it infinite} BP zone. 
According to Eq.(\ref{del1}), $\Delta$ vanishes for $k_4=k_2$ 
and the system is {\it always} in the BP zone. 
This is interesting, as then one 
can {\it never} `gain' by adding the extra path whatever be its (or other 
paths') characteristics.

\begin{figure}[tbh]
\centering
\rotatebox{270}{
\includegraphics[width=8cm,keepaspectratio]{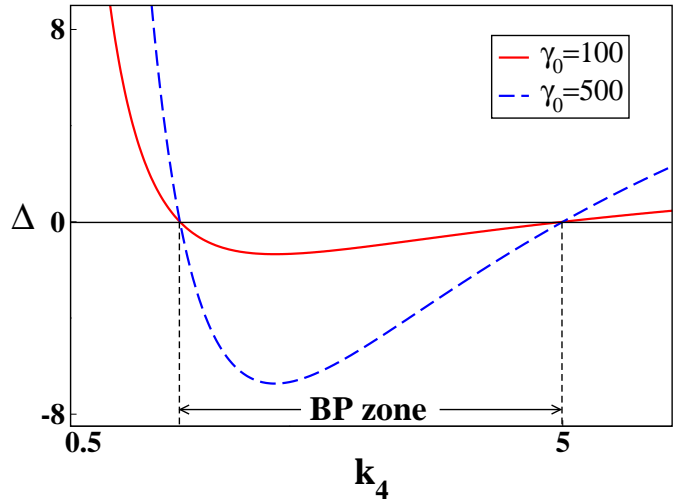}}
\caption{Variations of $\Delta$ as a function of 
$k_4$ for different $\gamma_0$ for the network in Scheme I. 
The values of the parameters are as follows: 
$k_1=1.0,\,k_2=5.0,\,k_3=1.5,\,k_6=10.0,\,k_5=5.0,\,k_{-5}=1.0$, 
all in ${\rm s^{-1}}$ and $\gamma_0$ is in ${\rm Ms^{-1}}$ unit.}
\label{f1}
\end{figure}

\subsection{Role of influx}

It is important to note also from Eqs.(\ref{J5ss})-(\ref{del1}) 
that, when other parameters are kept fixed, 
$\Delta$ is proportional to $\gamma_0$. Thus, 
the gain in adding the extra path is higher for a larger $\gamma_0$ 
in a non-BP zone ($\Delta>0$). But, so is the loss 
when the system resides in a BP zone. 
This feature is depicted in Fig.\ref{f1} by plotting $\Delta$ 
as a function $k_4$ for two values of $\gamma_0$ (in ${\rm Ms^{-1}}$). 
The BP zone is indicated explicitly in the figure, bounded 
by the two values of $k_4$ given by Eq.(\ref{del0}). 
Thus, for a network in the BP zone, the increase in load 
(after inclusion of extra path) can be extremely hazardous if it reaches 
or surpasses the inherent capacity. 
One can see from Fig.\ref{f1} that 
the variation of $\Delta$ is much sharper at smaller values 
of $k_4$. So, the transition from non-BP 
zone to BP zone can occur by a very minor alteration of $k_4$, 
particularly at higher $\gamma_0$. 
This sensitive dependence can lead even to a
network breakdown.

\subsection{Role of ${\rm C\to B}$ path}

\begin{figure}[tbh]
\centering
\rotatebox{270}{
\includegraphics[width=8cm,keepaspectratio]{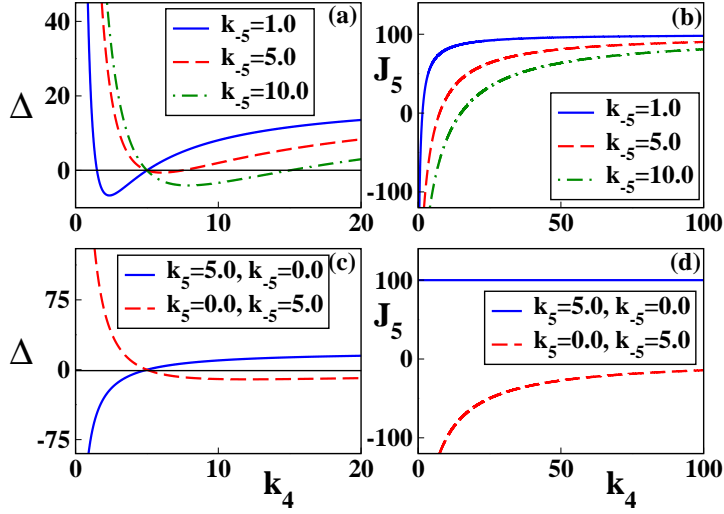}}
\caption{Variations of $\Delta$ and $J_5$ as a function of 
$k_4$ in Scheme I. 
The values of the relevant 
parameters are as follows:(a),(b) $k_1=1.0,\,k_2=5.0,\,k_3=1.5,
\,k_6=10.0,\,k_5=5.0$, all in ${\rm s^{-1}}$ and 
$\gamma_0=500\,{\rm Ms^{-1}}$.
(c),(d) $k_1=1.0,\,k_2=5.0,\,k_3=1.5,
\,k_6=10.0$, all in ${\rm s^{-1}}$ and $\gamma_0=500\,{\rm Ms^{-1}}$.}
\label{f2}
\end{figure}

We have plotted $\Delta$ in Fig.\ref{f2}(a) along with 
$J_5$ in Fig.\ref{f2}(b) for different values of 
$k_{-5}\,(>0)$. 
The relative positioning of the two points around which the 
sign changes of $\Delta$ occur are given by 
Eq.(\ref{del0}) again. The BP zones are thus {\it finite}. 
It follows from Eq.(\ref{J5ss}) that, 
at large values of $k_4$ (compared to the other rate constants), 
$J_5$ becomes independent of $k_4$ and then, using Eq.(\ref{del1}), 
one gets the limiting (positive) value of $\Delta$ as
\begin{equation}
\Delta=\frac{\gamma_0k_1k_5}{k_2(k_1+k_3)(k_2+k_5)}\,.
\label{delk4}
\end{equation}

\subsection{Specific cases}

In the rest of this section, we explore some specific 
cases. This will further improve our understanding of the 
variation of $\Delta$ and the significance of the BP zone.

\subsubsection{Case I: $k_5=0$ or $k_{-5}=0$}

This situation arises in an irreversible extra path. 
One finds here that $J_5$ cannot change sign and 
remains either positive ($k_{-5}=0$) or negative ($k_5=0$). 
Further, $J_5$ becomes independent of $k_4$ for $k_{-5}=0$ 
(see Fig.\ref{f2}(d)). 
So, $\Delta$ changes sign only once at $k_4=k_2$. 
For $k_5=0$, $\Delta$ is positive at 
low values of $k_4$, becomes zero at $k_4=k_2$ and then negative. 
This is shown in Fig.\ref{f2}(c). 
From Eq.(\ref{delk4}), it follows that 
$\Delta\to0$ at large $k_4$. Hence, if $k_4\ge k_2$, $\Delta\le0$ 
and the system supports an {\it infinite} BP zone . 
On the other hand, for $k_{-5}=0$, $\Delta$ is negative 
at small $k_4$ and $\Delta=0$ again at $k_4=k_2$ (see Fig.\ref{f2}(c)). 
So, the network remains in a {\it finite} BP zone for $k_4\le k_2$. 
Afterwards, $\Delta$ becomes positive and attains the 
value given in Eq.(\ref{delk4}) in the large-$k_4$ limit. 
Hence, the {\it nature of directionality} of the extra path 
strongly affects 
the emergence and sustenance of the BP zone. 
We mention that, when viewed as a function of $k_2$, 
the features 
simply get reversed. 
\begin{figure}[h!]
\centering
\rotatebox{270}{
\includegraphics[width=6.5cm,keepaspectratio]{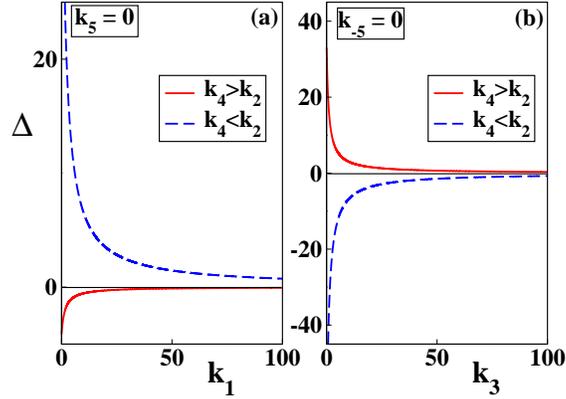}}
\caption{Variation of $\Delta$ (a) as a function of 
$k_1$ for $k_5=0$ and (b) as a function of $k_3$ for $k_{-5}=0$. 
Depending on the relative magnitudes of $k_2$ and $k_4$, 
the system can be entirely in the BP or in the non-BP zone. 
The values of the relevant 
parameters are as follows: (a)$k_2=5.0,\,k_3=1.5,\,k_6=10.0,\,
k_{-5}=1.0,\,k_4=2.0\,{\rm and},\,15.0$, all in ${\rm s^{-1}}$; 
(b) $k_1=1.0,\,k_2=5.0,\,k_6=10.0,\,
k_{5}=5.0,\,k_4=2.0\,{\rm and},\,15.0$, all in ${\rm s^{-1}}$. 
$\gamma_0=500\,{\rm Ms^{-1}}$ in both the cases.}
\label{f2a}
\end{figure}
Also, for vanishing $k_5$ or $k_{-5}$, 
$\Delta$ {\it cannot} change sign as a function of $k_1$ or 
$k_3$ (see Eq.(\ref{del0})). When these parameters are varied, 
the system remains entirely 
either in the BP or in the non-BP zone, depending on the relative 
magnitudes of $k_2,\,k_4$. For example, when $k_5=0$,  
$k_4\ge k_2$, the system remains {\it entirely} in the BP zone as 
a function of $k_1$. However, if $k_4<k_2$, then the BP 
zone {\it vanishes completely}. These properties are shown in 
Fig.\ref{f2a}(a). 
Similar plots are shown in Fig.\ref{f2a}(b) for $k_3$ at 
$k_{-5}=0$. 
For further details, we refer the reader to Table 1.

\subsubsection{Case II: $k_1k_5=k_3k_{-5}$}

\begin{figure}[tbh]
\centering
\rotatebox{270}{
\includegraphics[width=7cm,keepaspectratio]{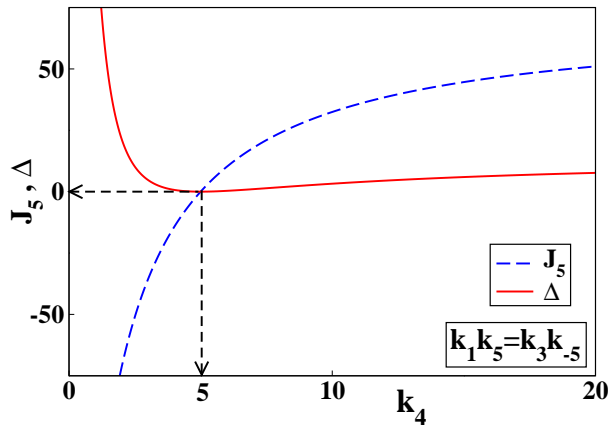}}
\caption{Variations of $\Delta$ and $J_5$ as a function of 
$k_4$ in Scheme I for the case $k_1k_5=k_3k_{-5}$. 
The values of the relevant 
parameters are as follows: $k_1=1.0,\,k_2=5.0,\,k_3=2.5,\,k_6=10.0,\,
k_5=5.0,\,k_{-5}=2.0$, all in ${\rm s^{-1}}$ and 
$\gamma_0=500\,{\rm Ms^{-1}}$.}
\label{f3}
\end{figure}

In this special case, the two points given in Eq.(\ref{del0}) 
merge to a single point, $k_4=k_2$. Using Eq.(\ref{J5ss}) and 
Eq.(\ref{del1}) along with the above condition, we get
\begin{equation}
\Delta=\frac{\gamma_0(k_4-k_2)^2k_1k_5}{k_2k_4(k_1+k_3)
(k_2(k_4+k_{-5})+k_4k_5)}.
\label{delEQ}
\end{equation} 
Hence, in this situation, $\Delta$ {\it cannot} be negative. 
The presence of the extra path decreases the load for all values 
of $k_4$ except at $k_4=k_2$ where $\Delta=0$ and the BP zone 
reduces to a point. These features are shown in Fig.\ref{f3} 
for specific values of the relevant parameters: 
$k_1=1.0,\,k_2=5.0,\,k_3=2.5,\,k_6=10.0,\,k_5=5.0,\,k_{-5}=2.0$, all in ${\rm s^{-1}}$ and $\gamma_0=500\,{\rm Ms^{-1}}$. 

\subsubsection{Case III: $k_1=k_4$, $k_3=k_2$}

This diagonally symmetric case resembles the networks 
commonly used to show the 
occurrence of BP \cite{Lep}. Here, the points where $\Delta$ 
is zero are given by 
\begin{equation}
k_4=k_2\left(\frac{k_{-5}}{k_5}\right)^{1/2}\,\,\,{\rm and}\,\,k_4=k_2.
\label{del01}
\end{equation}
Hence, $\Delta$ changes sign twice as a function of $k_4$. 
If either $k_5$ or $k_{-5}$ vanishes, then $\Delta$ can change 
sign once (at $k_4=k_2$). 
The two points in Eq.(\ref{del01}) merge at $k_5=k_{-5}$ whence 
$\Delta\ge0$. 
So, the main results remain unchanged in this symmetric case. 
A summary of all our findings on Scheme I can be found in 
Table 1. 

\begin{table}[h!]
\label{tab1}
\begin{center}
\caption{Summary of the results in the various cases of Scheme I 
regarding the appearance of BP zone as a function of different 
system parameters and the dependence on the nature of extra path. 
Here, $f=k_3k_{-5}/(k_1k_5)$, $k_4^*=fk_2,\,
k_4^{**}=k_2$, $k_2^*=k_4/f,\,k_2^{**}=k_4$, 
$k_1^*=k_2k_3k_{-5}/(k_4k_5)$, 
$k_3^*=k_1k_4k_5/(k_2k_{-5})$.}
\vskip .5cm 
\begin{tabular}{c|c|c}
\hline
\hline
Network parameter & Nature of extra path & BP zone condition\\
\hline
\hline
	& $k_{\pm5}>0$ & $k_4^*\le k_4\le k_4^{**}\,\,(f<1)$\\
	&	&	$k_4^{**}\le k_4\le k_4^{*}\,\,(f>1)$\\ 
\cline{2-3}
$k_4$	& $k_5=0$ & $k_4\ge k_4^{**}$\\	\cline{2-3}
	& $k_{-5}=0$ & $k_4\le k_4^{**}$\\ \cline{2-3}
	& $k_5/k_{-5}=k_3/k_1$ & $k_4=k_4^{**}$\\
\hline
	& $k_{\pm5}>0$ & $k_2^*\le k_2\le k_2^{**}\,\,(f>1)$\\
	&	&	$k_2^{**}\le k_2\le k_2^{*}\,\,(f<1)$\\ 
\cline{2-3}
$k_2$ 	& $k_5=0$ & $k_2\le k_2^{**}$\\	\cline{2-3}
	& $k_{-5}=0$ & $k_2\ge k_2^{**}$\\ \cline{2-3}
	& $k_5/k_{-5}=k_3/k_1$ & $k_2=k_2^{**}$\\
\hline
	& $k_{\pm5}>0$ & $k_1\le k_1^*\,\,(k_4\ge k_2)$\\ 
 	&	&	$k_1\ge k_1^*\,\,(k_4\le k_2)$\\ 
\cline{2-3}
$k_1$	& $k_5=0$ & $k_4\ge k_2$\\ \cline{2-3}
	& $k_{-5}=0$	& $k_4\le k_2$\\
\hline
	& $k_{\pm5}>0$ & $k_3\ge k_3^*\,\,(k_4>k_2)$\\ 
	&	&	$k_3\le k_3^*\,\,(k_4<k_2)$\\ 
\cline{2-3}
$k_3$	& $k_5=0$ & $k_4\ge k_2$\\ \cline{2-3}
	& $k_{-5}=0$	& $k_4\le k_2$\\
\hline
\hline
\end{tabular}
\end{center}
\end{table}

\section{Selective introduction of reversibility}

\begin{figure}[tbh]
\centering
\rotatebox{270}{
\includegraphics[width=7cm,keepaspectratio]{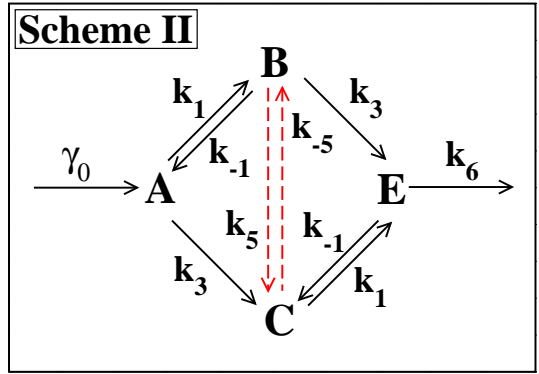}}
\caption{Schematic diagram of the symmetric CRN with 
some reversible steps. 
The extra path is shown by the dashed, red line.}
\label{sch2}
\end{figure}
The CRN, denoted as Scheme II, is shown in Fig.\ref{sch2}. 
This type of symmetric CRN was used recently 
in the same context \cite{Lep} where the reversible 
steps are taken to mimic traffic congestion. 
Due to the reversible steps, the fluxes $J_1(t)$ and $J_4(t)$ get 
modified as
\begin{equation}
J_1(t)=k_1a(t)-k_{-1}b(t);\,\,J_4(t)=k_1c(t)-k_{-1}e(t).
\label{Jnew}
\end{equation}
The other fluxes remain the same as in Scheme I, but 
with $k_2=k_3$. 
At SS, again we get $J_0=J_6$. 
The SS concentrations in this case are given by
\begin{equation}
b=\frac{\gamma_0k_1^2+(k_1+k_3)(k_{-1}+k_6)k_{-5}e}
{k_1^2(k_3+k_5)+k_1k_3(k_{-1}+k_3+k_5)+(k_1+k_3)k_3k_{-5}},
\label{bsnew}
\end{equation}
\begin{equation}
a=\frac{\gamma_0+k_{-1}b}{k_1+k_3};\,\,
k_1c=(k_{-1}+k_6)e-k_3b
\label{acsnew}
\end{equation}
where $e$ is given by Eq.(\ref{ess}). 
From Eq.(\ref{bsnew}), one gets
\begin{equation}
b^0=\frac{\gamma_0k_1}{k_3(k_1+k_{-1}+k_3)}.
\label{bs0}
\end{equation}

\subsection{Relation between load-difference and extra path flux}

At SS, the flux $J_5$ can be written as
$$J_5=J_4-J_3=J_6-J_2-J_3$$
$$=\gamma_0-k_3(a+b)$$
\begin{equation}
=\frac{\gamma_0k_1}{k_1+k_3}\left(1-\frac{b}{b^0}\right)
\label{J5new}
\end{equation}
where the last line is obtained by using Eqs.(\ref{acsnew})-(\ref{bs0}). 
The load-difference $\Delta$ 
turns out here as
$$\Delta=\left(\frac{k_1(k_1+k_{-1})-k_3^2}{k_1(k_1+k_3)}\right)
(b^0-b)$$
\begin{equation}
=\left(\frac{k_1(k_1+k_{-1})-k_3^2}{\gamma_0k_1^2}\right)
b^0J_5.
\label{delnew}
\end{equation}

\subsection{Exploring BP zones}

\begin{figure}[tbh]
\centering
\rotatebox{270}{
\includegraphics[width=7cm,keepaspectratio]{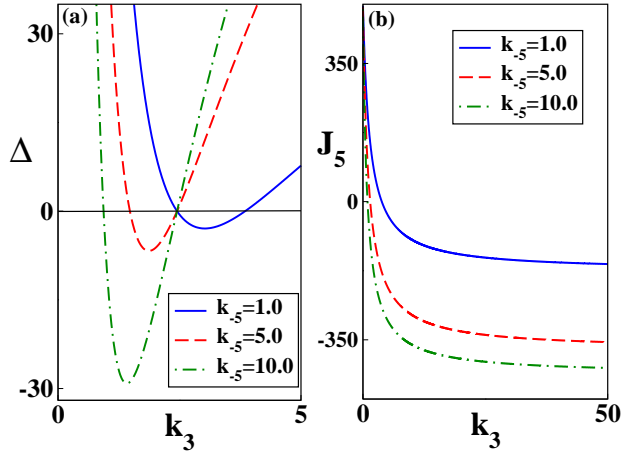}}
\caption{Variations of $\Delta$ and $J_5$ as a function of 
$k_3$ in Scheme II. 
The values of the relevant 
parameters are as follows: $k_1=2.0,\,k_{-1}=5.0,\,k_6=20.0,\,
k_5=5.0$, all in ${\rm s^{-1}}$ and $\gamma_0=500\,{\rm Ms^{-1}}$.}
\label{f5}
\end{figure}

We now calculate the points in the parameter space where 
$\Delta=0$ to find the BP zone(s). 
Putting Eq.(\ref{bsnew}) and Eq.(\ref{bs0}) into 
Eq.(\ref{delnew}) and setting the latter equal to zero, 
we get the equation
\begin{equation}
(k_1(k_1+k_{-1})-k_3^2)X=0.
\label{zero}
\end{equation}
Here 
\begin{equation}
X=P_1k_3^2+Q_1k_3+R_1
\label{x}
\end{equation}
with
\begin{equation}
P_1=\gamma_0k_{-5};\,Q_1=k_{-5}k_{-1}e(k_1+k_{-1}+k_3+k_6);\,
R_1=-\gamma_0k_5k_1^2.
\label{p1}
\end{equation}
If the variation of $\Delta$ is studied as a function of 
$k_3$, then from Eqs.(\ref{zero})-(\ref{x}), we get the two 
points where $\Delta$ becomes zero as
\begin{equation}
k_3=(k_1(k_1+k_{-1}))^{1/2}\,;\,\,
k_3=\frac{-Q_1+(Q_1^2-4P_1R_1)^{1/2}}{2P_1}.
\label{del0new}
\end{equation}
Therefore, $\Delta$ changes sign twice with $k_3$.  
At the second value of $k_3$ in Eq.(\ref{del0new}), $J_5=0$. 
We have plotted $\Delta$ and $J_5$ with $k_3$ in Fig.\ref{f5}. 

One can also get similar behavior as a function of other 
rate constants. For example, it follows from Eqs.(\ref{zero})-(\ref{x}) 
that $\Delta$ is zero at the following pair of values of $k_1$
\begin{equation}
k_1=\frac{-k_{-1}+(k_{-1}^2+4k_3^2)^{1/2}}{2}\,;\,\,
k_1=\frac{-Q_2-(Q_2^2-4P_2R_2)^{1/2}}{2P_2},
\label{del0new1}
\end{equation}
where 
\begin{equation}
P_2=-\gamma_0k_{5};\,Q_2=k_{-1}k_3k_{-5}e;\,
R_2=k_3k_{-5}(k_{-1}^2e+\gamma_0k_3+k_{-1}k_3e+k_{-1}\gamma_0).
\label{p2}
\end{equation}
Similarly, one obtains the equivalent points in terms 
of $k_{-1}$ as 
\begin{equation}
k_{-1}=\frac{k_3^2-k_1^2}{k_1}\,;\,\,
k_{-1}=\frac{-Q_3+(Q_3^2-4P_3R_3)^{1/2}}{2P_3},
\label{del0new2}
\end{equation}
where 
\begin{equation}
P_3=k_3k_{-5}e;\,Q_3=k_3k_{-5}(k_1e+k_3e+\gamma_0);\,
R_3=\gamma_0(k_{-5}k_3^2-k_{5}k_1^2).
\label{p3}
\end{equation} 
However, this last equation shows that, 
for $\Delta$ to change sign twice, one needs 
$R_3<0$ for a physically meaningful value of $k_{-1}$. 
It is easy to see that, for an irreversible extra path 
(vanishing $k_5$ or $k_{-5}$), 
$\Delta$ can change sign only once with the variation 
of any of the three parameters. The $\Delta=0$ points 
are given by the first values in 
Eq.(\ref{del0new}), Eq.(\ref{del0new1}) and Eq.(\ref{del0new2}), 
in the respective cases.

\section{Introduction of extra nodes}

\begin{figure}[tbh]
\centering
\rotatebox{270}{
\includegraphics[width=7cm,keepaspectratio]{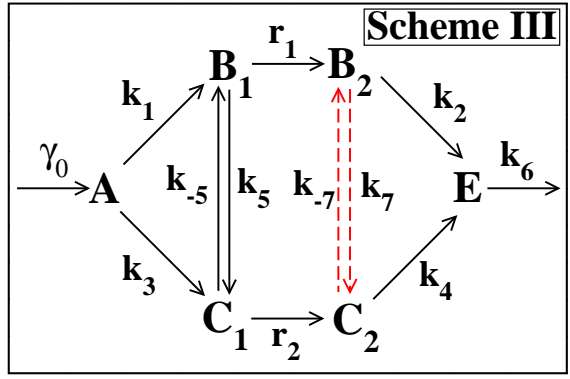}}
\caption{Schematic diagram of the extended, six-node CRN. 
The extra path is shown by the dashed, red line.}
\label{sch3}
\end{figure}

It is appropriate now to ask: how the basic features 
of the SS load will be affected in presence of additional 
nodes (species)? 
In this section, we try to answer this question. 
To do so, we consider an extended, six-node version of the 
CRN in Scheme I, denoted as Scheme III. 
The schematic is depicted in Fig.\ref{sch3}. 
To maintain the connection with the previous schemes, 
the route with rate constants $k_{\pm7}$ is taken 
as the extra path. This enables one to understand the role 
of the network structure in governing the SS load $Z$, 
particularly by comparing the results of this scheme with Scheme I. 
Following similar procedures as before, 
one obtains the SS concentrations as
\begin{equation}
a=\gamma_0/(k_1+k_3),\,e=\gamma_0/k_6,
\label{SS1}
\end{equation}
\begin{equation}
c_1=\frac{\gamma_0k_5+r_1k_3a}{r_1(r_2+k_{-5})+r_2k_5},\,
b_1=\frac{k_1a+k_{-5}c_1}{r_1+k_5},
\label{SS2}
\end{equation}
\begin{equation}
c_2=(\gamma_0k_7+k_2r_2c_1)/Y,\,b_2=(\gamma_0-k_4c_2)/k_2
\label{SS3}
\end{equation}
where 
\begin{equation}
Y=k_2(k_4+k_{-7})+k_4k_7.
\label{Y}
\end{equation}

Let us remind the reader that the variables at $k_{\pm7}=0$ 
are denoted by a zero superscript with $a^0=a,\,e^0=e,\,
c_1^0=c_1,\,b_1^0=b_1$ (see Eqs.(\ref{SS1})-(\ref{SS3})). 
From Eqs.(\ref{SS1})-(\ref{Y}), the effect of the extra path on $Z$ is 
determined by the following load-difference 
$$\Delta=Z^0-Z=
(b_1^0+c_1^0+b_2^0+c_2^0)-(b_1+c_1+b_2+c_2)$$
\begin{equation}
=\left(\frac{k_4-k_2}{Yk_2}\right)
\left(\gamma_0k_7-\frac{r_2(k_4k_7+k_2k_{-7})}{k_4}c_1\right).
\label{del3}
\end{equation}
It follows from Eq.(\ref{del3}) that $\Delta$ becomes zero 
at $k_4=k_2$ and also when 
\begin{equation}
c_1=\frac{\gamma_0k_4k_7}{r_2(k_4k_7+k_2k_{-7})}.
\label{cc1}
\end{equation}
Equating the two expressions of $c_1$ in Eq.(\ref{cc1}) and 
Eq.(\ref{SS2}), we obtain the required condition as
\begin{equation}
r_1k_4k_7(k_1r_2+k_1k_{-5}+k_3k_{-5})=
r_2k_2k_{-7}(r_1k_3+k_1k_5+k_3k_5).
\label{cond}
\end{equation}
So, as a function of $k_4$ or $k_2$, we again find that 
$\Delta$ can change sign twice. Also, for either $k_7=0$ or 
$k_{-7}=0$, $\Delta$ can change sign only once (see Eq.(\ref{cond})).  Thus, the qualitative behavior of $\Delta$ 
{\it remains the same} as 
that obtained via Scheme I or Scheme II.

\section{A biochemical application: Uric acid degradation in 
purine catabolism}

\begin{figure}[h!]
\centering
\rotatebox{270}{
\includegraphics[width=7cm,keepaspectratio]{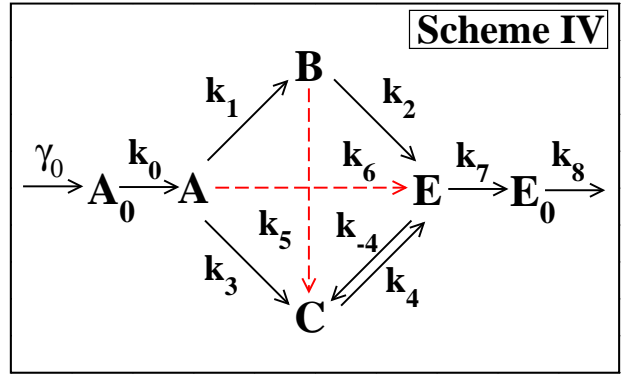}}
\caption{Schematic diagram of the uric acid degradation 
pathway, highly important in purine catabolism, 
based on Ref.\cite{Bov}. 
The species involved are: ${\rm A_0}:$ Uric acid, A: HIU, 
B: OHCU, C: $(R)$-Allantoin, E: $(S)$-Allantoin, 
${\rm E_0}:$ Allantoate. 
The extra paths (dashed, red lines) represent the spontaneous 
decomposition of intermediates A and B.}
\label{schex}
\end{figure}

The purine catabolism in many organisms involves 
the stereospecific breakdown of uric acid to produce allantoin 
as the key step \cite{Tip}. Notable exceptions are 
humans, birds, reptiles and some bacteria. 
This is the reason why accumulation of urate can lead 
to gout and renal stones in humans. The reaction pathway 
for oxidative decomposition of urate is shown schematically in 
Fig.\ref{schex}. 
This is based on the recent work 
by Bovigny {\it et al.} \cite{Bov}. 
The species consisting the CRN, denoted as Scheme IV, 
in Fig.\ref{schex} are: 
${\rm A_0}:$ Uric acid, A: 5-hydroxyisourate (HIU), 
B: 2-oxo-4-hydroxy-4-carboxy-5-ureidoimidazoline (OHCU), 
C: $(R)$-Allantoin, E: $(S)$-Allantoin, ${\rm E_0}:$ Allantoate. 
The enzymatic conversion of allantoin to allantoate (${\rm E_0}$) 
is stereospecific for the $(S)$-isomer (E). 
However, both the $(S)$- and the $(R)$-isomer are produced 
from the spontaneous (non-enzymatic) decomposition of HIU (A) and 
OHCU (B) and also via racemization. 
The details of the reaction steps and catalysts 
involved are given in Table 2.

\begin{table}[h!]
\label{tab2}
\begin{center}
\caption{Details of the CRN in Scheme IV representing 
uric acid degradation, based on Ref.\cite{Bov}. 
The abbreviations `sp' and `en' 
mean `spontaneous' and `enzymatic', respectively. 
The actual rate constants of 
enzymatic steps are denoted as $k'_i$ and those of the 
spontaneous decomposition steps as $k_i^0$.}
\begin{tabular}{c|c|c|c}
\hline
\hline
Reaction & Nature & Catalyst & Effective \\
	&	&	& rate constant\\
\hline
\hline
${\rm A_0\to A}$ & hydroxylation & Uricase (${\rm E_1}$) & 
$k_0=k'_0[{\rm E_1}]$ \\
\hline
${\rm A\to B}$ & hydrolysis & HIU hydrolase (${\rm E_2}$) & 
$k_1=k'_1[{\rm E_2}]$ \\
\hline
${\rm A\to C}$ & decomposition (sp) & $-$ & 
$k_3=k_3^0$ \\
\hline
${\rm B\to C}$ & decomposition (sp) & $-$ & 
$k_5=k_5^0$ \\
\hline
${\rm B\to E}$ & decarboxylation + & OHCU decarboxylase 
(${\rm E_3}$) &   \\
	& decomposition (sp) & $-$ & $k_2=k_2^0+k'_2[{\rm E_3}]$\\
\hline
${\rm A\to E}$ & decomposition (sp) & $-$ & 
$k_6=k_6^0$ \\
\hline
${\rm C\to E}$ & racemization (en)+ & Allantoin racemase 
(${\rm E_4}$) &	 \\
	& racemization (sp) & $-$ & $k_4=k_4^0+k'_4[{\rm E_4}]$\\
\hline
${\rm E\to E_0}$ & hydrolysis & Allantoinase (${\rm E_5}$) &  
$k_7=k'_7[{\rm E_5}]$ \\
\hline
\hline
\end{tabular}
\end{center}
\end{table}

In our context, the decomposition steps ${\rm A\to E}$ and 
${\rm B \to C}$ constitute the 
extra paths (shown by dashed, red lines in Fig.\ref{schex}). 
These paths result in the {\it indiscriminate} formation of 
species C and E from B and A, respectively. 
As the enzymatic degradations are stereospecific, the 
decomposition steps appear to be disadvantageous. 
Next, we will try to understand whether these pathways 
can have some {\it functional} role, {\it i.e.,} the system 
can use them for some kind of advantage. 
In what follows, 
the concentrations of different enzymes (and other 
species) 
involved in various steps of the network are taken as 
constants (acting as chemiostats) and 
included in the pseudo-first-order rate constants $k_i$. 
The variations of $k_i$ are thus 
naturally linked to the corresponding {\it parametric} variations 
in enzyme concentration. 

\subsection{Decomposition paths and BP zone}

The SS concentrations are determined following 
similar procedures as earlier. 
They are given by 
$$a_0=\frac{\gamma_0}{k_0},\,e=\frac{\gamma_0}{k_7},\,
e_0=\frac{\gamma_0}{k_8},$$
\begin{equation}
a=\frac{\gamma_0}{k_1+k_3+k_6},\,b=\frac{k_1a}{k_2+k_5},
\label{scon}
\end{equation}
$$c=\frac{k_3a+k_5b+k_{-4}e}{k_4}.$$ 
Using Eq.(\ref{scon}), the effect of the extra paths is determined 
by $\Delta$ as 
$$\Delta=Z(k_5=k_6=0)-Z=Z^{0}-Z$$
$$=(a^0+b^0+c^0)-(a+b+c)$$
\begin{equation}
=\frac{\gamma_0(s_1k_2^2+s_2k_2+s_3)}{k_2k_4(k_1+k_3)(k_2+k_5)s_4}
\label{delex}
\end{equation}
with the symbols having the same meanings as before. 
Here
$$s_1=k_6(k_3+k_4),\,s_2=k_5(k_4k_6+k_3k_6-k_1^2-k_1k_3)+k_1k_4k_6,$$
\begin{equation}
s_3=k_1k_4k_5s_4,\,s_4=(k_1+k_3+k_6).
\label{s123}
\end{equation}

\begin{figure}[h!]
\centering
\rotatebox{0}{
\includegraphics[width=8cm,keepaspectratio]{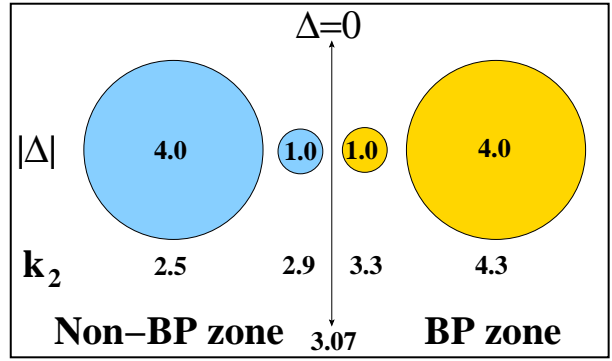}}
\caption{The variation in the magnitude of $\Delta$ 
as a function of $k_2$; $|\Delta|$ is proportional to the 
radii of the circles (written inside). 
The $\Delta=0$ points appear at $k_2=3.07$ ${\rm s^{-1}}$, 
shown in the figure, and at $k_2=1085.43$ ${\rm s^{-1}}$, 
for the set of parameters as follows: 
$k_1=10.0,\,k_3=1.0,\,k_4=3.0,\,k_{-4}=2.0,\,
k_5=4.0,\,k_6=0.1,\,k_7=1.5$, all in ${\rm s^{-1}}$ and 
$\gamma_0=100\,{\rm Ms^{-1}}$. }
\label{delp}
\end{figure}

Now, we investigate the possibility of BP zones in this 
network. As $s_1,s_3>0$, it 
follows from Eqs.(\ref{delex})-(\ref{s123}) that, $\Delta$ 
can't be zero for $s_2\ge0$ when viewed as a function of 
$k_2$. Actually, here $\Delta$ remains positive and 
there is {\it no scope} for BP zone to develop. 
Equation (\ref{s123}) suggests that this will occur for 
sufficiently large $k_5,\,k_6$ values compared to the other  
rate constants. Hence, the 
{\it indiscriminate decomposition pathways can play a 
functional role in reducing the SS load of the network}. 
If $s_2<0$, then 
it is possible for $\Delta$ to change sign twice with the 
variation of $k_2$ and {\it finite} BP zones appear. 
See Fig.\ref{delp}. 
The numerator in Eq.(\ref{delex}) is also quadratic in $k_1$. 
It can be easily checked that, depending on the relative magnitudes of 
$k_2$ and $k_4$, $\Delta$ can change sign once or twice with $k_1$. 
In this case, {\it infinite} BP zones can also emerge. 
Therefore, the variation in concentrations of different  
enzymes affects the system load in a distinct manner.

As there are two extra paths, another interesting feature is the 
effect of one in presence of the other. Let us first take 
the extra path BC. 
One obtains the effect of the BC path in presence of the 
AE path ($k_6>0$) as
\begin{equation}
\Delta_1=Z(k_5=0)-Z=\frac{(k_4-k_2)k_1k_5a}
{k_2k_4(k_2+k_5)}.
\label{del11}
\end{equation}
$\Delta_1$ becomes zero at $k_4=k_2$ and changes sign once 
against $k_2$ or $k_4$. {\it Infinite} BP zone appears when 
viewed as a function of $k_2$ whereas, the system 
supports a {\it finite} BP zone as a function of $k_4$. 
 
Next, we study the effect of the AE path 
in presence of the BC path ($k_5>0$). 
In this case, the load-difference becomes
$$\Delta_2=Z(k_6=0)-Z$$
\begin{equation}
=\left(\frac{\gamma_0k_6}{s_4(k_1+k_3)}\right)
\left[\left(1+\frac{k_3}{k_4}\right)+
\left(1+\frac{k_5}{k_4}\right)
\left(\frac{k_1}{k_2+k_5}\right)\right]>0.
\label{del22}
\end{equation}
Thus, in presence of the BC path, addition 
of the path AE {\it always} reduces the SS load 
and {\it no} BP zone exists. 
The above-mentioned features indicate that the SS load 
can play a major role, along with other factors, 
in governing the 
evolution of optimized reaction steps in living systems. 

\begin{figure}[h!]
\centering
\rotatebox{0}{
\includegraphics[width=5cm,keepaspectratio]{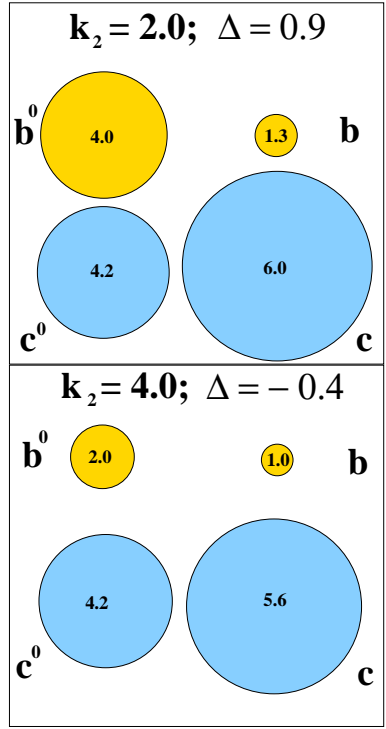}}
\caption{The SS concentrations of species B, C in absence 
and in presence of the extra paths (see Fig.\ref{schex}). The 
concentrations are proportional to the 
radii of the corresponding circles (written inside). 
In the top panel, the 
system is in the non-BP zone whereas, in the bottom panel, 
it is in the BP zone. The values of the relevant parameters 
are as follows: $k_1=10.0,\,k_3=1.0,\,k_4=3.0,\,k_{-4}=2.0,\,
k_5=4.0,\,k_6=0.1,\,k_7=1.5$, all in $s^{-1}$ and 
$\gamma_0=100\,{\rm Ms^{-1}}$. The SS concentration of 
species A alters little and hence not shown.}
\label{fpop}
\end{figure}

\subsection{The load distribution}

Here, we make a brief survey on the 
distribution of the load over the CRN, {\it i.e.,} on the 
individual SS concentrations. 
According to Eq.(\ref{scon}), only $a,\,b,\,c$ depend on the 
parameters of the extra paths. 
In absence the extra paths ($k_5=k_6=0$), 
$a,\,b$ increase whereas $c$ decreases. 
However, their {\it relative change in magnitudes} 
dictate the 
fate of $\Delta$. In Fig.\ref{fpop}, the changes in 
the SS concentrations due to the addition of the extra paths 
are shown at two values of $k_2$. 
For the parameters chosen, particularly at the 
small $k_6$ value, $a\approx a^0$ and hence not shown 
in the figure. Both in the top and bottom 
panels of Fig.\ref{fpop}, we have $b^0>b,\,c^0<c$. 
But, the system is in the BP zone only in the bottom 
panel. This shows that, focusing on a single species concentration 
can be misleading in detecting the BP zone. 

\subsection{The load dynamics}

\begin{figure}[h!]
\centering
\rotatebox{270}{
\includegraphics[width=8cm,keepaspectratio]{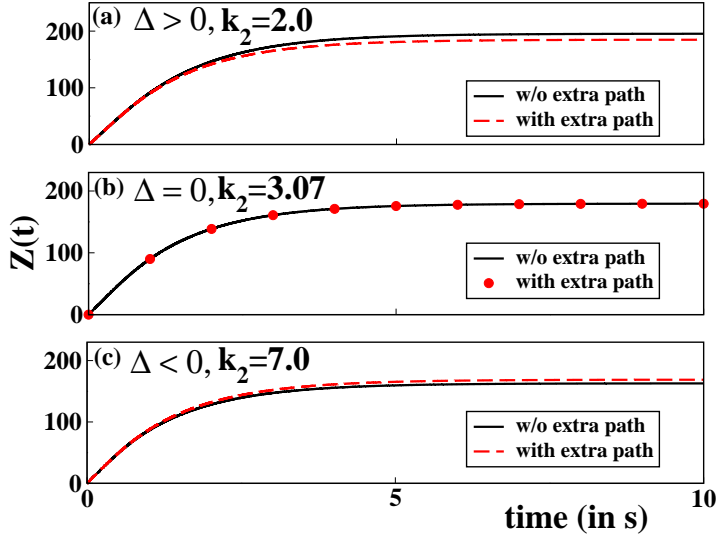}}
\caption{Time-evolution of $Z(t)$ in absence 
and in presence of the extra paths (see Fig.\ref{schex}). The three 
cases are: (a) $\Delta>0,\,k_2=2.0$ ${\rm s^{-1}}$, (b) $\Delta=0,\,
k_2=3.07$ ${\rm s^{-1}}$, (c) $\Delta<0,\,k_2=7.0$ ${\rm s^{-1}}$.
The values of the other parameters 
are as follows: $k_0=15.0,\,k_1=10.0,\,k_3=1.0,\,k_4=3.0,\,k_{-4}=2.0,\,
k_5=4.0,\,k_6=0.1,\,k_7=1.5,\,k_8=5.0$, all in ${\rm s^{-1}}$ and 
$\gamma_0=100\,{\rm Ms^{-1}}$.}
\label{fzt}
\end{figure}

Before concluding, we touch upon the dynamics of the load.  
The individual time-evolutions of the reacting species of Scheme IV 
are determined numerically. The load $Z(t)$ is plotted in 
Fig.\ref{fzt} as a function of time for three different cases: 
$\Delta>0,\,\Delta=0,\,\Delta<0$. The set of parameters are the same 
as in the previous two subsections. In all the cases, $\Delta$ shows 
a monotonic rise till the system reaches the SS. 
When the system is in the non-BP zone (Fig.\ref{fzt}a), $Z$ remains 
lower throughout if extra paths are present. Similarly, 
in the BP zone (Fig.\ref{fzt}c), $Z$ is greater in presence of extra 
paths during the whole reaction progression. Interestingly, for 
$\Delta=0$ (Fig.\ref{fzt}b), $Z$ follows virtually the same 
time-course with and without the extra paths. 
Thus, regarding the BP  zones, the load dynamics in our case is in conformity with the 
results obtained so far and does not provide any qualitatively new  information. 
However, for brevity, we refrain from a detailed analysis of the load  dynamics as a function of various system parameters. 

\section{Discussion and Conclusion}

In this study, we have investigated the effect of 
additional paths on the SS load of open CRNs. 
The problem is related to the handling of traffic 
flow in general and particularly the paradoxical case illustrated 
by Braess, where 
inclusion of extra routes increases the travel time. 
In an equivalent manner, we have found BP-like behavior 
in the SS load which, instead of dropping, 
can get raised or remains the same due to the presence of extra paths. 
The region of parameter space where this kind of behavior 
occurs is denoted as the BP zone. 
Taking some basic networks, we have explored the 
roles of the system parameters in governing the 
materialization of such zones. 
First of all, the load difference $\Delta$ is found to be 
proportional to the rate of influx $\gamma_0$. Thus, the BP-zone 
(as well as the non-BP zone) grows with $\gamma_0$. 
Investigating the effects of the paths 
forming network-edges, we have found an appealing aspect 
regarding the variation of the associated rate constants. 
For some of them ($k_2,\,k_4$ in Scheme I), the 
BP zone is {\it finite} as $\Delta$ changes sign twice. 
It can even reduce to a point. 
For the others ($k_1,\,k_3$ in Scheme I), 
$\Delta$ can change sign {\it only} once and then 
the BP zone can turn {\it infinite}. 
Making some steps reversible and symmetric (as in Scheme II) 
generates {\it finite} BP zones as a function 
of all the rate constants of the network edges. 
However, for an irreversible extra path, 
$\Delta$ can change sign only once in both the schemes. 
Then {\it finite} BP zones become {\it infinite} again. 
The basic features are shown to remain unchanged in an extended 
six-node version of the network (Scheme III).

The nature of the extra path is of immense importance regarding 
the existence of the BP zone. 
This is manifested in the intimate connection 
between $\Delta$ and $J_5$, the flux associated with 
the extra path (see Eq.(\ref{del1}) and Eq.(\ref{delnew})). 
For a reversible extra path, the 
BP zone {\it cannot} span the entire parameter ($k_i$) range 
although it can stretch infinitely. 
This is not always the case with an irreversible extra path. 
Depending on the direction of the extra path, 
the system can show two extreme behaviors. 
(i) It is {\it always} in the BP zone over the whole range 
of some parameter; 
(ii) It does {\it not} support any BP zone over the whole range 
of the same parameter. 
Between (i) and (ii), the kind of behavior actually shown 
by the system depends on the relative values of other 
parameters (case of $k_1,\,k_3$ in Scheme I, see Table 1). 
This kind of `switching' response of the system to the addition of 
extra path is highly interesting as well as of huge practical impact 
in the designing and planning of network geometries. 
We have also applied the methodology on the important 
biochemical network of uric acid degradation. 
From the detailed analyses of the corresponding CRN in an open 
system framework, we propose that some of the spontaneous 
decomposition steps of the intermediates can have 
{\it functional roles} in reducing the SS load. 
As the magnitude of the SS load is essential for the physical  sustainability of the reaction medium, this can be one of the 
major deciding factors in the evolution of various 
reaction mechanisms in living systems.

\section*{Acknowledgment}

K. Banerjee acknowledges the University Grants Commission 
(UGC), India for Dr. D. S. Kothari Fellowship. 
K. Bhattacharyya thanks CRNN, CU, for partial financial support.


\begin{thebibliography}{99}

\bibitem{Bert} L. von Bertalanffy, Science {\bf 111}, 23 (1950).

\bibitem{Horn} F. J. M. Horn and R. Jackson, Arch. Ration. Mech. Anal. 
{\bf 47}, 81 (1972).

\bibitem{Fein} M. Feinberg and F. J. M. Horn, Chem. Eng. Sci. 
{\bf 29}, 775 (1974).

\bibitem{Clr} B. L. Clarke, Adv. Chem. Phys. {\bf 43}, 1 (1980).


\bibitem{Schus} S. Schuster and R. Schuster, J. Math. Chem. 
{\bf 6}, 17 (1991).

\bibitem{Bhal} U. S. Bhalla, Prog. Biophys. Mol. Biol. {\bf 81},
45 (2003).

\bibitem{Wagn} G. P. Wagner, M. Pavlicev, and J. M. Cheverud, 
Nature Rev. Genetics {\bf 8}, 921 (2007).


\bibitem{Gold} A. Goldbeter, {\it Biochemical Oscillations and 
Cellular Rhythms. The Molecular Bases of Periodic and Chaotic 
Behavior}, (Cambridge Univ. Press, Cambridge, UK, 1996).

\bibitem{Edel} B. Edelstein, J. Theor. Biol. {\bf 29}, 5762 (1970). 

\bibitem{Fein1} M. Feinberg, Chem. Eng. Sci. {\bf 42} 2229 (1987).

\bibitem{Mer} M. Merrow and M. Brunner, FEBS Lett. {\bf 585}, 
1383 (2011).

\bibitem{Kelr} L. M. Kellershohn, Trends Biochem. Sci. {\bf 24}, 
418 (1999).

\bibitem{Jeo} H. Jeong, B. Tombor, R. Albert, Z. N. Oltvai, and 
A.-L. Barabasi, Nature {\bf 407}, 651 (2000).

\bibitem{Brw} E. Brown, R. Kass, and P. Mitra, Nature Neurosci. 
{\bf 7}, 456 (2004).

\bibitem{Fran} P. Francois and V. Hakim, Proc. Natl Acad. Sci. (USA) 
{\bf 101}, 580 (2004).

\bibitem{Slu} A. Slusarczyk, A. Lin, and R. Weiss, 
Nature Rev. Genet. {\bf 13}, 406 (2012). 

\bibitem{Meyr} D. Meyer, P. Neumann, E. Koers, H. Sjuts, S. Ludtke, 
G. M. Sheldrick, R. Ficner, and K. Tittmann, 
Proc. Natl. Acad. Sci. U.S.A. {\bf 109}, 10867 (2012).

\bibitem{Jac} D. Jacquemin, J. Zuniga, A. Requena, and 
J. P. Ceron-Carrasco, Acc. Chem. Res. (2014) DOI: 10.1021/ar500148c.


\bibitem{Katch} A. Katchalsky and Peter F. Curran, {\it 
Non-equilibrium thermodynamics in Biophysics}, (Harvard University Press,  Cambridge, MA, 1965.)

\bibitem{Prig3} G. Nicolis and I. Prigogine, {\it 
Self-organization in Non-equilibrium Systems (from Dissipative
Structures to order through Fluctuations)}, 
(John Wiley and Sons, New York, 1977.)

\bibitem{Mou} C. Y. Mou, J. -L. Luo, and G. Nicolis, 
J. Chem. Phys. {\bf 84}, 7011 (1986).

\bibitem{Ross} J. Ross and M. O. Vlad, Ann. Rev. Phys. Chem. 
{\bf 50}, 51 (1999).

\bibitem{gasprd1} P. Gaspard, J. Chem. Phys. {\bf 120}, 8898 (2004).

\bibitem{gasprd} D. Andrieux and P. Gaspard, J. Chem. Phys. {\bf 121}, 
6167 (2004).

\bibitem{seif1} T. Schmiedl, and U. Seifert, 
J. Chem. Phys. {\bf 126}, 044101 (2007). 


\bibitem{gasprd2} D. Andrieux and P. Gaspard, 
J. Chem. Phys. {\bf 130}, 014901 (2009).

\bibitem{Qan1} H. Qian, Annu. Rev. Phys. Chem. {\bf 58}, 113 (2007).

\bibitem{Qan2} M. Vellela and H. Qian, J. R. Soc. Interface 
{\bf 6}, 925 (2009).

\bibitem{Hop} J. J. Hopfield, Proc. Natl. Acad. Sci. U.S.A. {\bf 71}, 
4135 (1974).

\bibitem{TH} T. Hill, Proc. Natl. Acad. Sci. U.S.A {\bf 80}, 2922 (1983).

\bibitem{Jul} L. Jullien, A. Lemarchand, S. Charier, O. Ruel, and 
J.-B. Baudin, J. Phys. Chem. B {\bf 107}, 9905 (2003).

\bibitem{gaspnas} D. Andrieux and P. Gaspard, 
Proc. Natl. Acad. Sci. U.S.A. {\bf 105}, 9516 (2008).


\bibitem{Qan3} H. Qian and E. L. Elson, Biophys. Chem. {\bf 101}, 
565 (2002).

\bibitem{Min} W. Min, L. Jiang, J. Yu, S. C. Kou, H. Qian, and 
X. S. Xie, Nano Lett. {\bf 5}, 2373 (2005).

\bibitem{Banj} K. Banerjee, B. Das, and G. Gangopadhyay, 
J. Chem. Phys. {\bf 136}, 154502 (2012).

\bibitem{Pol} M. Polettini and M. Esposito, 
J. Chem. Phys. {\bf 141}, 024117 (2014). 


\bibitem{Tim} M. Timme and J. Casadiego, J. Phys. A: Math. Theor. 
{\bf 47}, 343001 (2014).

\bibitem{Bra} D. Braess, Unternehmensforschung {\bf 12}, 258 
(1968); English translation in: D. Braess, A. Nagurney, and 
T. Wakolbinger, Transportation Sci. {\bf 39}, 446 (2005).

\bibitem{Rou} T. Roughgarden, J. Comput. System Sci. {\bf 72}, 
922 (2006).

\bibitem{Pench} C. M. Penchina and L. J. Penchina, 
Am. J. Phys. {\bf 71}, 479 (2003).

\bibitem{Tim1} D. Witthaut and M. Timme, 
New J. Phys. {\bf 14}, 083036 (2012).

\bibitem{Pala} M. G. Pala, S. Baltazar, P. Liu, H. Sellier, 
B. Hackens, F. Martins, V. Bayot, X. Wallart, L. Desplanque, 
and S. Huant, Phys. Rev. Lett. {\bf 108}, 076802 (2012).

\bibitem{Lep} D. M. Lepore, C. Barratt, and P. M. Schwartz, 
J. Math. Chem. {\bf 49}, 356 (2011).



\bibitem{Mart} R. F. Martinez, M. Avalos, R. Babiano, P. Cintas, 
M. E. Light, J. L. Jimenez, and J. C. Palacios, 
Tetrahedron {\bf 70}, 2319 (2014).

\bibitem{Jon} R. A. Jones and A. Whitmore, ARKIVOC {\bf 11}, 114 (2007); 
DOI: http://dx.doi.org/10.3998/ark.5550190.0008.b10

\bibitem{Car} A. R. E. Carey, S. Eustace, R. A. M. O'Ferrall, and 
B. A. Murray, J. Chem. Soc. Perkin Trans. {\bf 2}, 2285 (1993) .

\bibitem{Nem} N. Nemeria, E. Binshtein, H. Patel, A. Balakrishnan, 
I. Vered, B. Shaanan,  Z. Barak, D. Chipman, and F. Jordan, 
Biochemistry {\bf 51}, 7940 (2012).


\bibitem{Tip} P. A. Tipton, Nat. Chem. Biol. {\bf 2}, 124 (2006).

\bibitem{Bov} C. Bovigny, M. T. Degiacomi, T. Lemmin, 
M. Dal Peraro, and M. Stenta, J. Phys. Chem. B 
{\bf 118}, 7457 (2014). 


\end{thebibliography}
\end{document}